\documentclass[aps,a4paper,superscriptaddress,preprintnumbers,showpacs,amsmath,amssymb]{revtex4}
\usepackage{graphicx}
\usepackage{dcolumn}
\usepackage{color}
\usepackage{latexsym,amsfonts}
\usepackage{textcomp}
\usepackage{bm}

\usepackage{ulem}

\def\beq{\begin{equation}}
\def\eeq{\end{equation}}

\begin{document}

\title{Effective Low--Energy Gravitational Potential\\ for Slow Fermions Coupled to Linearised Massive Gravity}
\author{A. N. Ivanov}\email{ivanov@kph.tuwien.ac.at}
\affiliation{Atominstitut, Technische Universit\"at Wien, Stadionallee
  2, A-1020 Wien, Austria}
\author{M. Pitschmann}\email{pitschmann@kph.tuwien.ac.at}
\affiliation{Atominstitut, Technische Universit\"at Wien, Stadionallee
  2, A-1020 Wien, Austria}
\author{M. Wellenzohn}\email{max.wellenzohn@gmail.com}
\affiliation{Atominstitut, Technische Universit\"at Wien, Stadionallee
  2, A-1020 Wien, Austria}

\date{\today}

\begin{abstract}
We analyse the Dirac equation for slow fermions coupled to 
linearised massive gravity above the Minkowski background and derive the
effective low--energy gravitational potential. The obtained results
can be used in terrestrial laboratories for the detection of
gravitational waves and fluxes of massive gravitons emitted by
cosmological objects. We also calculate the neutron spin precession within linearised massive gravity, which in principle can be measured by neutron interferometers.
\end{abstract}

 \maketitle

\section{Introduction}
\label{sec:introduction}

In this paper we analyse the low--energy reduction of Dirac fermions 
in interaction with linearised massive gravity.
The theory of massive gravity starts with the pioneering paper by
Fierz and Pauli \cite{Fierz1939}, in which they carried out the
analysis of the equations of motion of massive particles with spin
2. We shall call this graviton the Fierz--Pauli (FP) graviton. Very
nice surveys of the subsequent development of this theory and
gravitational theories beyond the cosmological standard model can be
found in the papers by Hinterbichler \cite{Hinterbichler2012} and by
Joyce {\it et al.}  \cite{Khoury2015}, respectively. As has been
pointed out by Brito, Cardoso and Pani \cite{Brito2013}, the
motivations to investigate massive gravity are conceptual
\cite{Hinterbichler2012} and practical, related to the analysis of the
influence of massive gravity on the dynamics of emission of
gravitational waves by pulsars
\cite{Damour1991,Finn2002,Chatziioannou2012} (see also
\cite{Brito2013}). The contribution of massive gravitons to
gravitational wave emission leads to a deformation of the
gravitational--wave signal during its journey from the source to
observer. As usual there are at least two ways for investigating
the deformation of the gravitational waveform caused by massive
gravitons. They are \cite{Brito2013}: i) a full non--linear simulation
and ii) a slow--motion expansion or a perturbative expansion around
some background. The later deals with a linear approximation of
non--linear massive gravity above a background \cite{Brito2013}, which
is valid at distances $r$ large compared to the so--called Vainshtein
radius $r_V = \sqrt[3]{M/M^2_{\rm Pl}m^2_g}$ \cite{Georgi2003} (see
also \cite{Hinterbichler2012} and \cite{Vainshtein2006}), i.e. $r > r_V$, where $M$ is the 
mass of the source producing gravitons, $M_{\rm Pl} = 1/\sqrt{8\pi G_N} =
2.435\times 10^{27}\,{\rm eV}$ is the reduced Planck mass defined in
terms of the Newtonian gravitational coupling constant $G_N$
\cite{PDG2014} and $m_g$ is the graviton mass.

The paper is organized as follows. In section \ref{sec:lgravity} we
follow \cite{Hinterbichler2012,Khoury2015} and derive the action of
linearised massive gravity above the Minkowski background by using the
St\"uckelberg trick to deal with the so--called van
Dam--Veltman--Zakharov (vDVZ) {\it discontinuity}
\cite{Dam1970,Zakharov1970}. In section \ref{sec:neutron} we analyse
the Dirac equation for slow fermions in linearised massive gravity
above the Minkowski background. In this section we follow the analysis
of the Dirac equation in curved spacetime as carried out by Kostelecky
\cite{Kostelecky2004}. However, in our analysis we
  neglect torsion. We calculate the spin connection and the Hamilton
operator in terms of the $\tilde{h}_{\mu\nu}$ field, which can be
identified with the field of the FP graviton including all shifts
related to the St\"uckelberg trick. In section \ref{sec:hamilton} we
investigate the transformation properties of the Hamilton operator
describing the interaction of fermions with a linearised massive
gravitational field. In section \ref{sec:foldy} we derive the
effective low--energy potential for slow fermions coupled to the
fields of linearised massive gravity above the Minkowski
background using Foldy--Wouthuysen transformations and write down the
corresponding Schr\"odinger--Pauli equation. In the restricted case of
a static diagonal metric, calculated in the weak field approximation,
we re-obtain our previous result for the effective low--energy
potential published in \cite{Ivanov2014}.  We want to
  stress that in the Dirac Hamilton operator as well as in the
  effective low-energy potential of the Schr\"odinger--Pauli equation
  the field degrees of freedom appear exclusively in terms of $\tilde
  h_{\mu\nu}$. In section \ref{sec:compare} we compare our results
  with those obtained by Gon\c{c}ales, Obukhov, and Shapiro
  \cite{Obukhov2007} as well as Quach \cite{Quach2015}, where similar
  calculations have been done for Dirac fermions in a
  gravitational--wave background. We calculate the neutron spin
  precession within linearised massive gravity. The phase--shift of
  the neutron wave function, caused by the neutron spin precession
  within linearised massive gravity, can in principle be measured by
  neutron interferometers \cite{Rauch2015}. In section
  \ref{sec:conclusion} we discuss the obtained results.

\section{Linearised massive gravity above Minkowski background}
\label{sec:lgravity}

We start with the action for linearised massive gravity
\cite{Hinterbichler2012,Khoury2015} coupled to Dirac fermions
with mass $m$. For the gravitational field we use the standard Einstein--Hilbert action
\begin{eqnarray}\label{eq:1}
S_{\rm EH} = \frac{M^2_{\rm Pl}}{2} \int d^4x\,\sqrt{- g}\,R\>,
\end{eqnarray}
where $M_{\rm Pl} = 1/\sqrt{8\pi G_N} = 2.435 \times 10^{27}\,{\rm
  eV}$ is the reduced Planck mass, $G_N$ the Newtonian
gravitational constant \cite{PDG2014}, $g_{\mu\nu}$ the metric
tensor with signature $(+1,-1,-1,-1)$, $g = {\rm det}\{g_{\mu\nu}\}$ the determinant of the
metric tensor, $R$ the Ricci scalar curvature given by
\cite{Feynman1995,Fliessbach2006,Rebhan2012}
\begin{eqnarray}\label{eq:2}
R = g^{\mu\lambda}{R^{\alpha}}_{\mu\alpha\lambda}\>,
\end{eqnarray}
and ${R^{\alpha}}_{\mu\nu\lambda}$ the Riemann--Christoffel
tensor defined by \cite{Feynman1995,Fliessbach2006,Rebhan2012}
\begin{eqnarray}\label{eq:3}
{R^{\alpha}}_{\mu\nu\lambda} = \frac{\partial
  {\Gamma^{\alpha}}_{\mu\nu}}{\partial x^{\lambda}} -
\frac{\partial{\Gamma^{\alpha}}_{\mu\lambda}}{\partial x^{\nu}} +
     {\Gamma^{\alpha}}_{\lambda\varphi}{\Gamma^{\varphi}}_{\mu\nu} -
     {\Gamma^{\alpha}}_{\nu\varphi}{\Gamma^{\varphi}}_{\mu\lambda}\>.
\end{eqnarray}
The affine connection ${\Gamma^{\alpha}}_{\mu\nu}$ is determined in
terms of the Christoffel symbols
\cite{Feynman1995,Fliessbach2006,Rebhan2012}
\begin{eqnarray}\label{eq:4}
{\Gamma^{\alpha}}_{\mu\nu} = \{{^{\alpha}}_{\mu\nu}\} =
\frac{1}{2}\,g^{\alpha\beta}\Big(\frac{\partial g_{\beta\mu}}{\partial
  x^{\nu}} + \frac{\partial g_{\beta\nu}}{\partial x^{\mu}} -
\frac{\partial g_{\mu\nu}}{\partial x^{\beta}}\Big)\>.
\end{eqnarray}
Integrating by parts and using the properties of the affine connection
and the metric tensor \cite{Rebhan2012a} we arrive at the following expression for the Einstein--Hilbert action
\begin{eqnarray}\label{eq:5}
S_{\rm EH} = \frac{M^2_{\rm Pl}}{2}\int d^4x\,\sqrt{- g}\,g^{\mu\nu}\Big(
{\Gamma^{\alpha}}_{\mu\nu}{\Gamma^{\beta}}_{\alpha\beta} - {\Gamma^{\alpha}}_{\mu\beta}{\Gamma^{\beta}}_{\alpha\nu}\Big)\>.
\end{eqnarray}
Next, we perform an expansion around a Minkowski background. For this, we set $g_{\mu\nu} = \eta_{\mu\nu} + h_{\mu\nu}$ and $g^{\nu\alpha} =
\eta^{\nu\alpha} - h^{\nu\alpha} + \mathcal O(h^2)$ such that $g_{\mu\nu}g^{\nu\alpha} =
    {\delta_{\mu}}^{\alpha}$, where $\eta_{\mu\nu}$ and
    $\eta^{\nu\alpha}$ are the Minkowski metric tensor and its inverse, respectively. To leading order we obtain
\begin{eqnarray}\label{eq:6}
\sqrt{- g}\,g^{\mu\nu}\Big(
{\Gamma^{\alpha}}_{\mu\nu}{\Gamma^{\beta}}_{\alpha\beta} - {\Gamma^{\alpha}}_{\mu\beta}{\Gamma^{\beta}}_{\alpha\nu}\Big) =
\frac{1}{2}\Big(\frac{1}{2}\frac{\partial h_{\mu\nu}}{\partial
  x_{\alpha}}\frac{\partial h^{\mu\nu}}{\partial x^{\alpha}} -
\frac{\partial h_{\nu\alpha}}{\partial x^{\mu}}\frac{\partial
  h^{\mu\alpha}}{\partial x_{\nu}} + \frac{\partial h_{\mu\nu}}{\partial x_{\mu}}\frac{\partial
  h}{\partial x_{\nu}} - \frac{1}{2}\frac{\partial h}{\partial
  x_{\alpha}}\frac{\partial h}{\partial x^{\alpha}}\Big)\>,
\end{eqnarray}
where $h = \eta^{\mu\nu}h_{\mu\nu}$. The structure of Eq.~(\ref{eq:6})
agrees well with Eq.~(6.83) of Ref.~\cite{Khoury2015}. Using
Eq.~(\ref{eq:6}) and adding the Pauli-Fierz mass term
\cite{Hinterbichler2012,Khoury2015}, we obtain the action for massive
gravity
\begin{eqnarray}\label{eq:7}
S_{\rm mgr} = \frac{M^2_{\rm Pl}}{4}\int
d^4x\,\Big(\frac{1}{2}\frac{\partial h_{\mu\nu}}{\partial
  x_{\alpha}}\frac{\partial h^{\mu\nu}}{\partial x^{\alpha}} -
\frac{\partial h_{\nu\alpha}}{\partial x^{\mu}}\frac{\partial
  h^{\mu\alpha}}{\partial x_{\nu}} + \frac{\partial
  h_{\mu\nu}}{\partial x_{\mu}}\frac{\partial h}{\partial x_{\nu}} -
\frac{1}{2}\frac{\partial h}{\partial x_{\alpha}}\frac{\partial
  h}{\partial x^{\alpha}} - \frac{1}{2}\,m^2_g\, \big(h_{\mu\nu}h^{\mu\nu} -
h^2\big)\Big)\>,
\end{eqnarray}
with graviton mass $m_g$. The main problem of this theory
is that it does not agree with the massless theory after performing the limit $m_g \to 0$, which is known as the
vDVZ discontinuity \cite{Dam1970,Zakharov1970}. It is useful to apply the so--called {\it St\"uckelberg
  trick}, related to the shift
\begin{eqnarray}\label{eq:8}
h_{\mu\nu} \to h_{\mu\nu} + \frac{\partial A_{\nu}}{\partial x^{\mu}}
+ \frac{\partial A_{\mu}}{\partial x^{\nu}}\>,
\end{eqnarray}
where $A_{\mu}$ is a new vector field
\cite{Hinterbichler2012,Khoury2015}. Such a trick
has been applied for the first time in the context of massive gravity in \cite{Georgi2003}. Since the shift Eq.~(\ref{eq:8})
is equivalent to an infinitesimal local shift in coordinates $x^\mu \to x^\mu - A^\mu(x)$, 
the terms corresponding to the Ricci scalar in Eq.~(\ref{eq:7}) are invariant under
the shift Eq.~(\ref{eq:8}). After performing this shift, the action of massive gravity
takes the form
\begin{eqnarray}\label{eq:9}
S_{\rm mgr} &=& \frac{M^2_{\rm Pl}}{4}\int
d^4x\,\Big\{\frac{1}{2}\frac{\partial h_{\mu\nu}}{\partial
  x_{\alpha}}\frac{\partial h^{\mu\nu}}{\partial x^{\alpha}} -
\frac{\partial h_{\nu\alpha}}{\partial x^{\mu}}\frac{\partial
  h^{\mu\alpha}}{\partial x_{\nu}} + \frac{\partial
  h_{\mu\nu}}{\partial x_{\mu}}\frac{\partial h}{\partial x_{\nu}} -
\frac{1}{2}\frac{\partial h}{\partial x_{\alpha}}\frac{\partial
  h}{\partial x^{\alpha}} - \frac{1}{2}\,m^2_g\,\big (h_{\mu\nu}h^{\mu\nu} -
h^2\big)\nonumber\\ &&+\> 2 m^2_g \Big[ - \frac{1}{4}\,F_{\mu\nu} F^{\mu\nu}
  - \Big(h^{\mu\nu}\frac{\partial A_{\nu}}{\partial x^{\mu}} - h\,
  \frac{\partial A_{\mu}}{\partial x_{\mu}}\Big) \Big]\Big\}\>.
\end{eqnarray}
The next step is to perform the shift
\begin{eqnarray}\label{eq:10}
A_{\mu} \to A_{\mu} - \frac{\partial \varphi}{\partial x^{\mu}}\>.
\end{eqnarray}
This gives
\begin{eqnarray}\label{eq:11}
S_{\rm mgr} &=& \frac{M^2_{\rm Pl}}{4}\int
d^4x\,\Big\{\frac{1}{2}\frac{\partial h_{\mu\nu}}{\partial
  x_{\alpha}}\frac{\partial h^{\mu\nu}}{\partial x^{\alpha}} -
\frac{\partial h_{\nu\alpha}}{\partial x^{\mu}}\frac{\partial
  h^{\mu\alpha}}{\partial x_{\nu}} + \frac{\partial
  h_{\mu\nu}}{\partial x_{\mu}}\frac{\partial h}{\partial x_{\nu}} -
\frac{1}{2}\frac{\partial h}{\partial x_{\alpha}}\frac{\partial
  h}{\partial x^{\alpha}} - \frac{1}{2}\,m^2_g\,\big (h_{\mu\nu}h^{\mu\nu} -
h^2\big)\nonumber\\ &&+\> 2 m^2_g \Big[ - \frac{1}{4}\,F_{\mu\nu} F^{\mu\nu}
- \Big(h^{\mu\nu}\frac{\partial A_{\nu}}{\partial x^{\mu}} - h\,
\frac{\partial A_{\mu}}{\partial x_{\mu}}\Big) - \Big(\frac{\partial
  h^{\mu\nu}}{\partial x^{\mu}}\frac{\partial\varphi}{\partial
  x^{\nu}} - \frac{\partial h}{\partial x_{\alpha}} \frac{\partial
  \varphi}{\partial x^{\alpha}}\Big)\Big]\Big\}\>,
\end{eqnarray}
where we have integrated by parts in the last term. Following
\cite{Hinterbichler2012,Khoury2015}, we make rescalings of the fields
\begin{eqnarray}\label{eq:12}
h_{\mu\nu} \to \frac{2}{M_{\rm Pl}}\,h_{\mu\nu}\>,\qquad A_{\mu} \to \frac{1}{M_{\rm Pl}}\frac{\sqrt{2}}{m_g}\,A_{\mu}\>,\qquad \varphi \to \frac{1}{M_{\rm Pl}}\frac{2d}{m^2_g}\,\varphi\>,
\end{eqnarray}
where $d$ is a dimensionless parameter to be determined later and all fields have the same dimension $[h_{\mu}] = [A_{\mu}] =
[\varphi] = {\rm eV}$. Hence, we
arrive at the action
\begin{eqnarray}\label{eq:13}
S_{\rm mgr} &=& \int d^4x\,\Big\{\frac{1}{2}\frac{\partial
  h_{\mu\nu}}{\partial x_{\alpha}}\frac{\partial h^{\mu\nu}}{\partial
  x^{\alpha}} - \frac{\partial h_{\nu\alpha}}{\partial
  x^{\mu}}\frac{\partial h^{\mu\alpha}}{\partial x_{\nu}} +
\frac{\partial h_{\mu\nu}}{\partial x_{\mu}}\frac{\partial h}{\partial
  x_{\nu}} - \frac{1}{2}\frac{\partial h}{\partial
  x_{\alpha}}\frac{\partial h}{\partial x^{\alpha}} -
\frac{1}{2}\,m^2_g \,\big(h_{\mu\nu}h^{\mu\nu} - h^2\big)\nonumber\\ &&-\>
\frac{1}{4}\,F_{\mu\nu} F^{\mu\nu} - \sqrt{2}
m_g\,\Big(h^{\mu\nu}\frac{\partial A_{\nu}}{\partial x^{\mu}} - h\,
\frac{\partial A_{\mu}}{\partial x_{\mu}}\Big) - 2d
\Big(\frac{\partial h^{\mu\nu}}{\partial
  x^{\mu}}\frac{\partial\varphi}{\partial x^{\nu}} - \frac{\partial
  h}{\partial x_{\alpha}} \frac{\partial \varphi}{\partial
  x^{\alpha}}\Big)\Big\}\>.
\end{eqnarray}
Following \cite{Hinterbichler2012,Khoury2015}, we perform the shift
\begin{eqnarray}\label{eq:14}
h_{\mu\nu} \to h_{\mu\nu} + d\,\varphi\,\eta_{\mu\nu}\>.
\end{eqnarray}
Substituting Eq.~(\ref{eq:14}) into Eq.~(\ref{eq:13}) we obtain the
following action
\begin{eqnarray}\label{eq:15}
S_{\rm mgr} &=& \int d^4x\,\Big\{\frac{1}{2}\frac{\partial
  h_{\mu\nu}}{\partial x_{\alpha}}\frac{\partial h^{\mu\nu}}{\partial
  x^{\alpha}} - \frac{\partial h_{\nu\alpha}}{\partial
  x^{\mu}}\frac{\partial h^{\mu\alpha}}{\partial x_{\nu}} +
\frac{\partial h_{\mu\nu}}{\partial x_{\mu}}\frac{\partial h}{\partial
  x_{\nu}} - \frac{1}{2}\frac{\partial h}{\partial
  x_{\alpha}}\frac{\partial h}{\partial x^{\alpha}} -
\frac{1}{2}\,m^2_g \,\big(h_{\mu\nu}h^{\mu\nu} - h^2\big)\nonumber\\ &&-\>
\frac{1}{4}\,F_{\mu\nu} F^{\mu\nu} - \sqrt{2}
m_g\,\Big(h^{\mu\nu}\frac{\partial A_{\nu}}{\partial x^{\mu}} - h\,
\frac{\partial A_{\mu}}{\partial x_{\mu}}\Big) + 3 d^2 \frac{\partial
  \varphi}{\partial x_{\alpha}}\frac{\partial \varphi}{\partial x^{\alpha}}
+ 3dm^2_g\,\varphi\,\big(h + 2 d\, \varphi\big) + 3\sqrt{2}d m_g\,
\varphi\,\frac{\partial A_{\mu}}{\partial x_{\mu}}\Big\}\>.
\end{eqnarray}
The problem of action Eq.~(\ref{eq:15}) is that the mass term of
the scalar particle $\varphi$ has incorrect sign. According to
\cite{Hinterbichler2012}, this problem can be remedied via 
gauge fixing, which furthermore leads to the diagonalisation of
action Eq.~(\ref{eq:15}). Hence, we add the terms
\begin{eqnarray}\label{eq:16}
S_{\rm gf1} = \int d^4x\,\Big(\frac{\partial
  h_{\nu\alpha}}{\partial x_{\nu}} - \frac{1}{2}\frac{\partial
  h}{\partial x^{\alpha}} - \frac{1}{\sqrt{2}}\,m_g A_{\alpha}\Big)\Big(\frac{\partial
  h^{\mu\alpha}}{\partial x^{\mu}} - \frac{1}{2}\frac{\partial
  h}{\partial x_{\alpha}} - \frac{1}{\sqrt{2}}\,m_g A^{\alpha}\Big)\>,
\end{eqnarray}
and 
\begin{eqnarray}\label{eq:17}
S_{\rm gf2} = -\frac{1}{2}\int d^4x\,\Big(\frac{\partial
  A_{\mu}}{\partial x_{\mu}} + \frac{1}{\sqrt{2}}\,m_g h + 3d\sqrt{2}m_g\,\varphi\Big)^2\>,
\end{eqnarray}
fixing the gauge this way. Hence, we arrive at the diagonal action 
\begin{eqnarray}\label{eq:18}
S_{\rm mgr} &=& \int
d^4x\,\Big\{\Big(\frac{1}{2}\frac{\partial h_{\mu\nu}}{\partial
  x_{\alpha}}\frac{\partial h^{\mu\nu}}{\partial x^{\alpha}} -
\frac{1}{2}\,m^2_g h_{\mu\nu}h^{\mu\nu}\Big) + \Big(-
\frac{1}{4}\frac{\partial h}{\partial x_{\alpha}}\frac{\partial
  h}{\partial x^{\alpha}} + \frac{1}{4}\,m^2_g\,h^2\Big)\nonumber\\
&&+\> \Big(- \frac{1}{2}\frac{\partial A_{\alpha}}{\partial x_{\mu}}
\frac{\partial A^{\alpha}}{\partial x^{\mu}} + \frac{1}{2}\,m^2_g
A_{\alpha}A^{\alpha}\Big) + 3d^2\Big(\frac{\partial \varphi}{\partial
  x_{\alpha}}\frac{\partial \varphi}{\partial x^{\alpha}} - m^2_g\,
\varphi^2\Big)\Big\}\>.
\end{eqnarray}
In order to obtain the correct kinetic term of the
massive scalar field we set the parameter $3 d^2 = 1/2$, finally arriving at
\begin{eqnarray}\label{eq:19}
S_{\rm mgr} &=& \int d^4x\,\Big\{\Big(\frac{1}{2}\frac{\partial
  h_{\mu\nu}}{\partial x_{\alpha}}\frac{\partial h^{\mu\nu}}{\partial
  x^{\alpha}} - \frac{1}{2}\,m^2_g h_{\mu\nu}h^{\mu\nu}\Big) + \Big(-
\frac{1}{4}\frac{\partial h}{\partial x_{\alpha}}\frac{\partial
  h}{\partial x^{\alpha}} + \frac{1}{4}\,m^2_g\,h^2\Big)\nonumber\\ &&+\>
\Big(- \frac{1}{2}\frac{\partial A_{\alpha}}{\partial x_{\mu}}
\frac{\partial A^{\alpha}}{\partial x^{\mu}} + \frac{1}{2}\,m^2_g
A_{\alpha}A^{\alpha}\Big) + \Big(\frac{1}{2}\frac{\partial
  \varphi}{\partial x_{\alpha}}\frac{\partial \varphi}{\partial x^{\alpha}}
- \frac{1}{2}\,m^2_g\, \varphi^2\Big)\Big\}\>.
\end{eqnarray}
It is important to note that all fields $h_{\mu\nu}$, $A_{\mu}$ and
$\varphi$ have masses equal to $m_g$. Now we may proceed to analyse the
gravitational interaction of slow fermions within linearised
massive gravity.

\section{Dirac equation for slow fermions in massive gravity}
\label{sec:neutron}

For the analysis of the Dirac equation for slow fermions we start with the action
\begin{eqnarray}\label{eq:20}
{\rm S}_{\psi} = \int d^4x\,\sqrt{-
  g}\,\Big(\frac{i}{2}\,\bar{\psi}(x) \gamma^{\mu}(x)\!
\stackrel{\leftrightarrow}{D}_{\mu}\!\psi(x) - m\bar{\psi}(x)\psi(x)\Big)\>,
\end{eqnarray}
where $\bar{\psi}(x) \gamma^{\mu}(x)\!\!
\stackrel{\leftrightarrow}{D}_{\mu}\!\!\psi(x) = \bar{\psi}(x)
\gamma^{\mu}(x) D_{\mu}\psi(x) - (\bar{\psi}(x)\stackrel{\leftarrow}{D}_{\mu})\gamma^{\mu}(x)
\psi(x)$, $m$ is the fermion mass, $\gamma^{\mu}(x)$ are the
Dirac matrices in curved spacetime satisfying the anticommutation
relations
\begin{eqnarray}\label{eq:21}
\gamma^{\mu}(x)\gamma^{\nu}(x) +
\gamma^{\nu}(x)\gamma^{\mu}(x) = 2 g^{\mu\nu}(x)\>,
\end{eqnarray}
and $D_{\mu}$ is the covariant derivative. For the
definition of the Dirac matrices $\gamma^{\mu}(x)$ and
covariant derivative $D_{\mu}$ we follow
\cite{Kostelecky2004} (see also
Eq.~(\ref{eq:28})). However, unlike Kostelecky \cite{Kostelecky2004}
our analysis is carried out for vanishing torsion. 
In order to derive the low--energy approximation 
of the Dirac equation in curved spacetime, we introduce a set of vierbein fields
$e^{\hat{\alpha}}_{\mu}(x)$ at each spacetime point $x$
defined by
\begin{eqnarray}\label{eq:22}
dx^{\hat{\alpha}} = e^{\hat{\alpha}}_{\mu}(x) dx^{\mu}\>.
\end{eqnarray}
In terms of these vierbein fields slow fermions couple to the
gravitational field. The vierbein fields $e^{\hat{\alpha}}_{\mu}(x)$
are related to the metric tensor $g_{\mu\nu}(x)$ by
\begin{eqnarray}\label{eq:23}
ds^2 = \eta_{\hat{\alpha}\hat{\beta}}
\,dx^{\hat{\alpha}} dx^{\hat{\beta}} =
\eta_{\hat{\alpha}\hat{\beta}}\,\big(e^{\hat{\alpha}}_{\mu}(x)dx^{\mu}\big)
    \big(e^{\hat{\beta}}_{\nu}(x)dx^{\nu}\big) =
    \big(\eta_{\hat{\alpha}\hat{\beta}}\,e^{\hat{\alpha}}_{\mu}(x)
      e^{\hat{\beta}}_{\nu}(x)\big) dx^{\mu}dx^{\nu} =
    g_{\mu\nu}(x)dx^{\mu}dx^{\nu}\>,
\end{eqnarray}
where $\eta_{\hat{\alpha}\hat{\beta}}$ is the metric tensor in 
Minkowski spacetime. This gives
\begin{eqnarray}\label{eq:24}
g_{\mu\nu}(x) =
\eta_{\hat{\alpha}\hat{\beta}}\,e^{\hat{\alpha}}_{\mu}(x)
e^{\hat{\beta}}_{\nu}(x) = e^{\hat{\alpha}}_{\mu}(x)
e_{\hat{\alpha}\nu}(x)\>.
\end{eqnarray}
Hence, the vierbein fields may be viewed as the square root of the metric
tensor $g_{\mu\nu}(x)$ in the sense of a matrix equation
\cite{Fischbach1981}. Inverting relation Eq.~(\ref{eq:22}) we
obtain
\begin{eqnarray}\label{eq:25}
\eta_{\hat{\alpha}\hat{\beta}}= g_{\mu\nu}(x)
e^{\mu}_{\hat{\alpha}}(x) e^{\nu}_{\hat{\beta}}(x) = e^{\mu}_{\hat{\alpha}}(x)
e_{\hat{\beta}\mu}(x)\>.
\end{eqnarray}
The following important relations hold
\begin{eqnarray}\label{eq:26}
e^{\mu}_{\hat{\alpha}}(x) e^{\hat{\beta}}_{\mu}(x) &=&
\delta^{\hat{\beta}}_{\hat{\alpha}}\>,\nonumber\\ e^{\mu}_{\hat{\alpha}}(x)
e^{\hat{\alpha}}_{\nu}(x) &=&
\delta^{\mu}_{\nu}\>,
\end{eqnarray}
which are useful for the derivation of the Dirac equation and
calculation of the Dirac Hamilton operator. In terms of the vierbein fields the Dirac
matrices $\gamma^{\mu}(x)$ are given by
\begin{eqnarray}\label{eq:27}
\gamma^{\mu}(x) =
e^{\mu}_{\hat{\alpha}}(x) \gamma^{\hat{\alpha}}\>,
\end{eqnarray}
where $\gamma^{\hat{\alpha}}$ are the standard Dirac matrices in Minkowski
spacetime \cite{Itzykson1980}. The covariant derivative $D_{\mu}$ we
define as \cite{Kostelecky2004}
\begin{eqnarray}\label{eq:28}
D_{\mu}\psi(x) = \partial_{\mu}\psi(x) -
\Gamma_{\mu}(x)\psi(x)\>,\qquad (\bar{\psi}(x)\stackrel{\leftarrow}{D}_{\mu}) =
\partial_{\mu}\bar{\psi}(x) -
\bar{\psi}(x)\gamma^{\hat{0}}\Gamma^{\dagger}_{\mu}(x)\gamma^{\hat{0}}\>,
\end{eqnarray}
where $\Gamma_{\mu}(x)$ is the spin affine connection, which can be expressed in
terms of the spin connection
$\omega_{\mu\hat{\alpha}\hat{\beta}}(x)$ 
\cite{Kostelecky2004}
\begin{eqnarray}\label{eq:29}
\Gamma_{\mu}(x) =
\frac{i}{4}\,\omega_{\mu\hat{\alpha}\hat{\beta}}
\sigma^{\hat{\alpha}\hat{\beta}}\>,
\end{eqnarray}
where $\sigma^{\hat{\alpha}\hat{\beta}} = (i/2)
(\gamma^{\hat{\alpha}}\gamma^{\hat{\beta}} -
\gamma^{\hat{\beta}}\gamma^{\hat{\alpha}})$ and
$\gamma^{\hat{0}}\Gamma^{\dagger}_{\mu}(x)\gamma^{\hat{0}} = -
  \Gamma_{\mu}(x)$. The spin connection
  $\omega_{\mu\hat{\alpha}\hat{\beta}}(x)$ is related to the vierbein
  fields and the affine connection as follows \cite{Kostelecky2004}
\begin{eqnarray}\label{eq:30}
\omega_{\mu\hat{\alpha}\hat{\beta}}(x) = -
\eta_{\hat{\alpha}\hat{\varphi}}\Big(\partial_{\mu}e^{\hat{\varphi}}_{\nu}(x)
- {\Gamma^{\alpha}}_{\mu\nu}(x) e^{\hat{\varphi}}_{\alpha}(x)\Big)
e^{\nu}_{\hat{\beta}}(x)\>,
\end{eqnarray}
where the affine connection ${\Gamma^{\alpha}}_{\mu\nu}(x)$ is defined in terms of the
Christoffel symbols only, i.e.  ${\Gamma^{\alpha}}_{\mu\nu}(x) =  \Big\{\begin{array}{c}
  \alpha \\
  \mu\nu \\
  \end{array}\Big\}$
\cite{Feynman1995,Fliessbach2006,Rebhan2012}.
The integrand of the action Eq.~(\ref{eq:20}) is hermitian. Integrating
by parts we transcribe the action Eq.~(\ref{eq:20}) into the form
\begin{eqnarray}\label{eq:31}
{\rm S}_{\psi} &=& \int d^4x\,\sqrt{-
  g}\,\Big\{i\,e^{\mu}_{\hat{\lambda}}(x)\bar{\psi}(x)
\gamma^{\hat{\lambda}}\partial_{\mu}\psi(x) -
\frac{i}{2}\,e^{\mu}_{\hat{\lambda}}(x) \bar{\psi}(x)
\gamma^{\hat{\lambda}} \Gamma_{\mu}(x) \psi(x) +
\frac{i}{2}\,e^{\mu}_{\hat{\lambda}}(x) \bar{\psi}(x) \gamma^{\hat{0}}
\Gamma^{\dagger}_{\mu}(x) \gamma^{\hat{0}} \gamma^{\hat{\lambda}}\psi(x)\nonumber\\
&& +\> \frac{i}{2}\frac{1}{\sqrt{ - g}}\frac{\partial}{\partial
  x^{\mu}}\Big(\sqrt{-
  g}\,e^{\mu}_{\hat{\lambda}}(x)\Big)\,\bar{\psi}(x)\gamma^{\hat{\lambda}}
\psi(x) - m \bar{\psi}(x) \psi(x)\Big\}\>.
\end{eqnarray}
Rewriting the fourth term in the integrand 
\begin{eqnarray}\label{eq:32}
\frac{i}{2}\frac{1}{\sqrt{- g}}\frac{\partial
}{\partial x^{\mu}}(\sqrt{-
  g}\,e^{\mu}_{\hat{\lambda}}(x))
\gamma^{\hat{\lambda}} = - \frac{i}{2}\,
\omega_{\mu\hat{\alpha}\hat{\beta}}(x) e^{\mu}_{\hat{\lambda}}(x)
\eta^{\hat{\lambda}\hat{\beta}}\gamma^{\hat{\alpha}}\>,
\end{eqnarray}
the action Eq.~(\ref{eq:31}) can be transformed into the form
\begin{eqnarray}\label{eq:33}
{\rm S}_{\psi} &=& \int d^4x\,\sqrt{-
  g}\,\Big\{i\,e^{\mu}_{\hat{\lambda}}(x)\bar{\psi}(x)
\gamma^{\hat{\lambda}}\partial_{\mu}\psi(x) -
\frac{i}{2}\,e^{\mu}_{\hat{\lambda}}(x) \bar{\psi}(x)
\gamma^{\hat{\lambda}} \Gamma_{\mu}(x) \psi(x) +
\frac{i}{2}\,e^{\mu}_{\hat{\lambda}}(x) \bar{\psi}(x)
\gamma^{\hat{0}} \Gamma^{\dagger}_{\mu}(x) \gamma^{\hat{0}}
\gamma^{\hat{\lambda}}\psi(x)\nonumber\\ && -\>
\frac{i}{2}\,\omega_{\mu\hat{\alpha}\hat{\beta}}(x) 
e^{\mu}_{\hat{\lambda}}(x)\,\eta^{\hat{\lambda}\hat{\beta}}
\bar{\psi}(x)\gamma^{\hat{\alpha}}\psi(x) - m \bar{\psi}(x)
\psi(x)\Big\}\>,
\end{eqnarray}
respectively
\begin{eqnarray}\label{eq:34}
{\rm S}_{\psi} &=& \int d^4x\,\sqrt{-
  g}\,\Big\{i\,e^{\mu}_{\hat{\lambda}}(x)\bar{\psi}(x)
\gamma^{\hat{\lambda}}\partial_{\mu}\psi(x) -
\frac{i}{2}\,\omega_{\mu\hat{\alpha}\hat{\beta}}(x)
e^{\mu}_{\hat{\lambda}}(x)\,
\bar{\psi}(x)\Big(\eta^{\hat{\lambda}\hat{\beta}}
\gamma^{\hat{\alpha}} +
\frac{i}{4}\,\{\sigma^{\hat{\alpha}\hat{\beta}},\gamma^{\hat{\lambda}}\}
\Big)\psi(x) - m \bar{\psi}(x) \psi(x)\Big\}\>. \nonumber\\
\end{eqnarray}
Using the following relations for Dirac matrices
\begin{eqnarray}\label{eq:35}
\gamma^{\hat{\lambda}}\sigma^{\hat{\alpha}\hat{\beta}} &=&
i\,(\eta^{\hat{\lambda}\hat{\alpha}}\,\gamma^{\hat{\beta}} -
\eta^{\hat{\beta}\hat{\lambda}}\,\gamma^{\hat{\alpha}}) -
\varepsilon^{\hat{\lambda}\hat{\alpha}\hat{\beta}\hat{\rho}}\gamma_{\hat{\rho}}
\gamma^5\>,\nonumber\\ \sigma^{\hat{\alpha}\hat{\beta}}\gamma^{\hat{\lambda}}
&=& i\,(\eta^{\hat{\beta}\hat{\lambda}}\,\gamma^{\hat{\alpha}} -
\eta^{\hat{\lambda}\hat{\alpha}}\,\gamma^{\hat{\beta}}) -
\varepsilon^{\hat{\lambda}\hat{\alpha}\hat{\beta}\hat{\rho}}\gamma_{\hat{\rho}}
\gamma^5\>,\nonumber\\ \{\sigma^{\hat{\alpha}\hat{\beta}},
\gamma^{\hat{\lambda}}\} &=& - 2\,
\varepsilon^{\hat{\lambda}\hat{\alpha}\hat{\beta}\hat{\rho}}\gamma_{\hat{\rho}}
\gamma^5\>,
\end{eqnarray}
where $\varepsilon^{\hat{\lambda}\hat{\alpha}\hat{\beta}\hat{\rho}}$
is the Levi--Civita tensor ($\varepsilon^{\hat{0}\hat{1}\hat{2}\hat{3}}
= +1$), we get
\begin{eqnarray}\label{eq:36}
{\rm S}_{\psi} &=& \int d^4x\,\sqrt{-
  g}\,\Big\{i\,e^{\mu}_{\hat{\lambda}}(x)\bar{\psi}(x)
\gamma^{\hat{\lambda}}\partial_{\mu}\psi(x) -
\frac{i}{2}\,\omega_{\mu\hat{\alpha}\hat{\beta}}(x)
e^{\mu}_{\hat{\lambda}}(x)\,
\bar{\psi}(x)\Big(\eta^{\hat{\lambda}\hat{\beta}}
\gamma^{\hat{\alpha}} -
\frac{i}{2}\,\varepsilon^{\hat{\lambda}\hat{\alpha}\hat{\beta}\hat{\rho}}\gamma_{\hat{\rho}}
\gamma^5 \Big)\psi(x) - m \bar{\psi}(x) \psi(x)\Big\}\>. \nonumber\\
\end{eqnarray}
After all shifts and rescalings, carried out in section
\ref{sec:lgravity}, the expansion of the metric tensor $g_{\mu\nu}$ above Minkowski spacetime acquires the form
\begin{eqnarray}\label{eq:37}
g_{\mu\nu} = \eta_{\mu\nu} + 2 \tilde{h}_{\mu\nu}\>,
\end{eqnarray}
where $\tilde{h}_{\mu\nu}$ is defined by ($d = \pm 1/\sqrt{6}$)
\begin{eqnarray}\label{eq:38}
\tilde{h}_{\mu\nu} = \frac{1}{M_{\rm Pl}}\Big(h_{\mu\nu} +
\frac{1}{\sqrt{2} m_g}\,\Big(\frac{\partial A_{\nu}}{\partial x^{\mu}}
+ \frac{\partial A_{\mu}}{\partial x^{\nu}}\Big) -
\frac{2d}{m^2_g}\frac{\partial^2 \varphi}{\partial x^{\mu}\partial
  x^{\nu}} + d\,\varphi\,\eta_{\mu\nu}\Big)\>.
\end{eqnarray}
The vierbein fields are given by
\begin{eqnarray}\label{eq:39}
e^{\hat{\alpha}}_{\mu} = \delta^{\hat{\alpha}}_{\mu} +
\tilde{h}^{\hat{\alpha}}_{\mu}\>,\qquad e^{\nu}_{\hat{\beta}} =
\delta^{\nu}_{\hat{\beta}} - \tilde{h}^{\nu}_{\hat{\beta}}\>.
\end{eqnarray}
We would like to note that in comparison with the
  metric decomposition $g_{\mu\nu} = \eta_{\mu\nu} + h_{\mu\nu}$ as
  used in section \ref{sec:lgravity} below Eq.~(\ref{eq:5}), in the
  decomposition $g_{\mu\nu} = \eta_{\mu\nu} + 2\tilde{h}_{\mu\nu}$ the
  weak gravitational field enters with a factor 2. This is done in
  order to use the vierbein fields Eq.~(\ref{eq:39}) without
  additional factors of $1/2$ in front of
  $\tilde{h}^{\hat{\alpha}}_{\mu}$ and
  $\tilde{h}^{\nu}_{\hat{\beta}}$.  To linear order
  the vierbein fields Eq.~(\ref{eq:39}) satisfy the relations in
  Eq.~(\ref{eq:26}). The spin connection
  $\omega_{\mu\hat{\alpha}\hat{\beta}}$, calculated to linear order,
  reads
\begin{eqnarray}\label{eq:40}
\omega_{\mu\hat{\alpha}\hat{\beta}} = \frac{\partial \tilde{h}_{
    \hat{\alpha}\mu}}{\partial x^{\hat{\beta}}} - \frac{\partial
  \tilde{h}_{\hat{\beta}\mu}}{\partial x^{\hat{\alpha}}}\>.
\end{eqnarray}
Plugging it into Eq.~(\ref{eq:36}) we arrive at the action
\begin{eqnarray}\label{eq:41}
{\rm S}_{\psi} = \int d^4x\,\sqrt{-
  g}\,\Big\{i\,e^{\mu}_{\hat{\lambda}}(x)\bar{\psi}(x)
\gamma^{\hat{\lambda}}\partial_{\mu}\psi(x) -
\frac{i}{2}\,\Big(\frac{\partial \tilde{h}_{
    \hat{\alpha}\mu}}{\partial x_{\mu}} - \frac{\partial
  \tilde{h}}{\partial x^{\hat{\alpha}}}\Big)\, \bar{\psi}(x)
\gamma^{\hat{\alpha}} \psi(x) - m \bar{\psi}(x)
\psi(x)\Big\}\>.
\end{eqnarray}
The Dirac equation in the standard form is 
\begin{eqnarray}\label{eq:42}
i\,\frac{\partial \psi}{\partial t} = {\rm H}\,\psi\>,
\end{eqnarray}
where ${\rm H} = {\rm H}_0 + \delta {\rm H}$ is the Dirac Hamilton
operator for fermions with mass $m$; ${\rm H}_0 = \gamma^0m
- i\gamma^0 \vec{\gamma}\cdot \vec{\nabla}$ is the Hamilton operator
for free fermions and $\delta {\rm H}$ describes
the interaction of fermions with linearised
massive gravity
\begin{eqnarray}\label{eq:43}
\delta {\rm H} &=& \tilde{h}^0_0\gamma^0\,m - i\gamma^0\gamma^j
\tilde{h}^0_0\frac{\partial}{\partial x^j} + i\gamma^0\gamma^j
\tilde{h}^k_j\frac{\partial}{\partial x^k} +
\frac{i}{2}\,\gamma^0\gamma^{\alpha}\,\Big(\frac{\partial
  \tilde{h}^{\mu}_{ \alpha}}{\partial x^{\mu}} - \frac{\partial
  \tilde{h}}{\partial x^{\alpha}}\Big) + i
  \gamma^0\gamma^j \tilde{h}^0_j \frac{\partial}{\partial t} + i
  \tilde{h}^j_0\frac{\partial}{\partial x^j}\>,
\end{eqnarray}
where we have dropped the hat on the indices and have kept only the linear order contributions of
$\tilde{h}_{\mu\nu}$. 

\section{Non--unitary transformations of Wave Functions and Hamilton Operator}
\label{sec:hamilton}

 It is well--known that the Hamilton operator of a Dirac particle in
 curved spacetime is not Hermitian
 \cite{Fischbach1981,Obukhov2001}. The corresponding Hermitian
 Hamilton operator may be obtained by means of a non--unitary
 transformation of its wave function \cite{Obukhov2014} (see also
 \cite{Fischbach1981,Obukhov2001} and \cite{Ivanov2014})
\begin{eqnarray}\label{eq:44}
\psi(x) = \big(\sqrt{- g}\,e^0_{\hat{0}}\big)^{-1/2}\,\psi'(x)\>.
\end{eqnarray}
The Hermitian Hamilton operator ${\rm H}'$ is related to the
non--Hermitian ${\rm H}$ by
\begin{eqnarray}\label{eq:45}
{\rm H}' = \big(\sqrt{- g}\,e^0_{\hat{0}}\big)^{+1/2}\,{\rm H}\big(\sqrt{-
    g}\,e^0_{\hat{0}}\big)^{-1/2} - i\,\big(\sqrt{- g}\,e^0_{\hat{0}}\big)^{+1/2}\frac{\partial}{\partial t}\big(\sqrt{- g}\,e^0_{\hat{0}}\big)^{-1/2}\>.
\end{eqnarray}
Keeping only linear contributions in the
$\tilde{h}_{\mu\nu}$--expansion and using $\sqrt{- g}\,e^0_{\hat{0}} =
1 + \big(\tilde{h} - \tilde{h}^0_0\big)$ we have to perform the non--unitary
transformation of the wave function of the Dirac particle
\begin{eqnarray}\label{eq:46}
\psi(x) =  \Big(1 - \frac{1}{2}\,\big(\tilde{h} -
\tilde{h}^0_0\big)\Big)\,\psi'(x)\>,
\end{eqnarray}
and arrive at the Hamilton operator
\begin{eqnarray}\label{eq:47}
\delta {\rm H}' = \delta {\rm H} - \frac{1}{2}\,\big[{\rm H}_0,\big(\tilde{h}
  - \tilde{h}^0_0\big)\big] + \frac{i}{2}\frac{\partial }{\partial
  t}\big(\tilde{h} - \tilde{h}^0_0\big)\>.
\end{eqnarray}
After the evaluation of the commutator
\begin{eqnarray}\label{eq:48}
- \frac{1}{2}\,\big[{\rm H}_0,\big(\tilde{h} - \tilde{h}^0_0\big)\big] =
\frac{i}{2}\,\gamma^0 \gamma^j\frac{\partial}{\partial
  x^j}\big(\tilde{h} - \tilde{h}^0_0\big)\>,
\end{eqnarray}
we obtain
\begin{eqnarray}\label{eq:49}
 \delta {\rm H}' = \tilde{h}^0_0\gamma^0\,m - i\gamma^0\gamma^j
 \tilde{h}^0_0\frac{\partial}{\partial x^j} + i\gamma^0\gamma^j
 \tilde{h}^k_j\frac{\partial}{\partial x^k} +
 \frac{i}{2}\,\gamma^0\gamma^j\,\Big(\frac{\partial
   \tilde{h}^{\mu}_j}{\partial x^{\mu}} - \frac{\partial
   \tilde{h}^0_0}{\partial x^j}\Big) +
 \frac{i}{2}\,\Big(\frac{\partial \tilde{h}^{\mu}_0}{\partial
   x^{\mu}} - \frac{\partial \tilde{h}^0_0}{\partial
   t}\Big) + i
  \gamma^0\gamma^j  \tilde{h}^0_j \frac{\partial}{\partial t} + i
   \tilde{h}^j_0\frac{\partial}{\partial x^j}\>. \nonumber\\
\end{eqnarray}
The Hamilton operator Eq.~(\ref{eq:49}) may also be transcribed into
the form
\begin{eqnarray}\label{eq:50}
 \delta {\rm H}' &=& \tilde{h}^0_0\gamma^0\,m - i\gamma^0\gamma^j
 \tilde{h}^0_0\frac{\partial}{\partial x^j} -
 \frac{i}{2}\,\gamma^0\gamma^j\frac{\partial \tilde{h}^0_0
 }{\partial x^j} + i\gamma^0\gamma^j
 \tilde{h}^k_j\frac{\partial}{\partial x^k} +
 \frac{i}{2}\,\gamma^0\gamma^j\frac{\partial \tilde{h}^k_j}{\partial
   x^k} + \frac{i}{2}\,\gamma^0\gamma^j\frac{\partial
   \tilde{h}^0_j}{\partial t} +
\frac{i}{2}\frac{\partial
   \tilde{h}^j_0}{\partial x^j}\nonumber\\ && +\>i
   \gamma^0\gamma^j \tilde{h}^0_j \frac{\partial}{\partial t} + i
   \tilde{h}^j_0\frac{\partial}{\partial x^j}\>.
\end{eqnarray}
One can show that the Hamilton operator Eq.~(\ref{eq:50}) is
hermitian. Introducing the notation
\begin{eqnarray}\label{eq:51}
  Q^\mu_\nu = \tilde{h}^0_0 \delta^\mu_\nu - \tilde{h}^\mu_\nu\>,\qquad Q_\nu :=
  \frac{\partial Q^\mu_\nu}{\partial x^\mu} = \frac{\partial \tilde{h}^0_0}{\partial x^\nu} - \frac{\partial
    \tilde{h}^\mu_\nu}{\partial x^\mu}\>,
\end{eqnarray}
allows to transcribe the Hamilton operator Eq.~(\ref{eq:50}) into the form
\begin{eqnarray}\label{eq:52}
 \delta {\rm H}' = \tilde{h}^0_0\gamma^0\,m - i\gamma^0\gamma^j
 Q^k_j\frac{\partial}{\partial x^k} - \frac{i}{2}\,\gamma^0\gamma^j
 Q_j  +
 \frac{i}{2}\frac{\partial
   \tilde{h}^j_0}{\partial x^j} + i \gamma^0\gamma^j \tilde{h}^0_j
   \frac{\partial}{\partial t} + i
   \tilde{h}^j_0\frac{\partial}{\partial x^j}\>,
\end{eqnarray}
which is convenient for the derivation of the effective low--energy
gravitational potential.

\section{Foldy--Wouthuysen transformations and low--energy 
approximation of the Dirac equation}
\label{sec:foldy}

For the derivation of the low--energy approximation of the Dirac
equation Eq.~(\ref{eq:49}) with the Hamilton operator ${\rm H}' = {\rm
  H}_0 + \delta {\rm H}'$, where $\delta {\rm H}'$ is given by
Eq.~(\ref{eq:52}), we employ the Foldy--Wouthuysen (FW) transformation
\cite{Foldy1950}. The aim of the FW transformation is obtain the rigorous low-energy 
limit by using unitary transformations \cite{Foldy1950}. First, we obtain
\begin{eqnarray}\label{eq:53}
{\rm H}_1 &=& e^{\,+ i S_1}\,{\rm H}'\,e^{\,-i S_1} - i\,e^{\,i
  S_1}\frac{\partial}{\partial t}e^{\,-i S_1} \nonumber\\
  &=& {\rm H}' -
\frac{\partial S_1}{\partial t} + i\Big[S_1,{\rm H}' -
  \frac{1}{2}\frac{\partial S_1}{\partial t}\Big] +
\frac{i^2}{2}\,\Big[S_1,\Big[S_1,{\rm H}' -
    \frac{1}{3}\frac{\partial S_1}{\partial t}\Big]\Big] + \ldots\>.
\end{eqnarray}
Following \cite{Foldy1950}, we take the
operator $S_1$ in the following form 
\begin{eqnarray}\label{eq:54}
S_1 &=& - \frac{i}{2m}\,\gamma^0\Big\{ - i
\gamma^0\gamma^j\frac{\partial }{\partial x^j} - i\gamma^0\gamma^j
Q^k_j\frac{\partial}{\partial x^k} - \frac{i}{2}\,\gamma^0\gamma^j
Q_j + i \gamma^0\gamma^j \tilde{h}^0_j
  \frac{\partial}{\partial t}\Big\}\nonumber\\ &=&-
\frac{1}{2m}\,\gamma^j\frac{\partial }{\partial x^j} -
\frac{1}{2m}\,\gamma^j Q^k_j\frac{\partial}{\partial x^k} -
\frac{1}{4m}\,\gamma^j Q_j + \frac{1}{2m}\, \gamma^j
  \tilde{h}^0_j \frac{\partial}{\partial t}\>,
\end{eqnarray}
The time derivative of $S_1$ and the commutators in Eq.~(\ref{eq:53})
are equal to
\begin{eqnarray}\label{eq:55}
\frac{\partial S_1}{\partial t} &=& -
\frac{1}{2m}\,\gamma^j\frac{\partial Q^k_j}{\partial
  t}\frac{\partial}{\partial x^k} - \frac{1}{4m}\,\gamma^j
\frac{\partial Q_j}{\partial t} + \frac{1}{2m}\,
  \gamma^j\frac{\partial \tilde{h}^0_j}{\partial
    t}\frac{\partial}{\partial t}\>,
\end{eqnarray}
respectively
\begin{eqnarray}\label{eq:56}
i\Big[S_1,{\rm H}' - \frac{1}{2}\frac{\partial
    S_1}{\partial t}\Big]
&=&i\,\gamma^0\gamma^j\frac{\partial}{\partial x^j} +
i\,\gamma^0\gamma^j Q^k_j\frac{\partial}{\partial x^k} +
\frac{i}{2}\,\gamma^0\gamma^j Q_j - i \gamma^0\gamma^j
\tilde{h}^0_j \frac{\partial}{\partial t} -
\gamma^0\,\frac{1}{m}\,\Delta\nonumber\\ &&+\>
\frac{i}{2}\,\gamma^0\gamma^j\frac{\partial
    \tilde{h}^0_0}{\partial x^j} + i\,\gamma^0 \gamma^j \tilde{h}^0_0
  \frac{\partial}{\partial x^j}\nonumber\\ &&+\>\gamma^0\,\frac{2}{m}\,
Q^k_j\frac{\partial^2}{\partial x^k \partial x_j} +
\gamma^0\,\frac{1}{m}\frac{\partial Q^k_j}{\partial
  x_j}\frac{\partial}{\partial x^k} +
\gamma^0\,\frac{i}{m}\,\varepsilon^{jkm} \Sigma_m \frac{\partial
  Q^{\ell}_k}{\partial x^j}\frac{\partial}{\partial
  x^{\ell}}\nonumber\\ &&+\> \gamma^0\,\frac{1}{m}\,Q_j
\frac{\partial}{\partial x_j} + \gamma^0\,\frac{1}{2m}\frac{\partial
  Q_j}{\partial x_j} +
\gamma^0\,\frac{i}{2m}\,\varepsilon^{jkm}\Sigma_m\frac{\partial
  Q_k}{\partial x^j},\nonumber\\ &&-\>\gamma^0\,
  \frac{2}{m}\, \tilde{h}^0_j \frac{\partial^2}{\partial x_j \partial t}
  - \gamma^0\,\frac{1}{m}\frac{\partial \tilde{h}^0_j}{\partial
    x_j}\frac{\partial}{\partial t} -
  \gamma^0\,\frac{i}{m}\,\varepsilon^{jkm}\Sigma_m \frac{\partial
    \tilde{h}^0_k}{\partial x^j}\frac{\partial}{\partial
    t}\nonumber\\ &&+\>
  \frac{1}{2m}\,\gamma^j\frac{\partial \tilde{h}^0_k}{\partial
    x^j}\frac{\partial}{\partial x_k} +
  \frac{1}{4m}\,\gamma^j\frac{\partial^2 \tilde{h}^0_k}{\partial x^j
    \partial x_k}\>,
\end{eqnarray}
and finally
\begin{eqnarray}\label{eq:57}
 \frac{i^2}{2}\,\Big[S_1,\Big[S_1,{\rm H}' - \frac{1}{3}\frac{\partial
       S_1}{\partial t}\Big]\Big]&=&\gamma^0\,\frac{1}{2m}\,\Delta +
 \gamma^0\,\frac{1}{2m}\,\vec{\nabla}\tilde{h}^0_0\cdot \vec{\nabla} +
 \gamma^0\,\frac{1}{8m}\,\Delta \tilde{h}^0_0 +
 \gamma^0\,\frac{1}{2m}\,\tilde{h}^0_0 \Delta +
 \gamma^0\,\frac{i}{4m}\,\vec{\Sigma}\cdot
 (\vec{\nabla}\tilde{h}^0_0 \times \vec{\nabla}\,) \nonumber\\ &&-\>
 \gamma^0\,\frac{1}{m}\,Q^k_j\frac{\partial^2}{\partial x_j \partial
   x^k} - \gamma^0\,\frac{1}{2m}\frac{\partial Q^k_j}{\partial
   x_j}\frac{\partial}{\partial x^k} -
 \gamma^0\,\frac{i}{2m}\,\varepsilon^{jkm}\Sigma_m \frac{\partial
   Q^{\ell}_k}{\partial x^j}\frac{\partial}{\partial
   x^{\ell}}\nonumber\\ &&-\>
 \gamma^0\,\frac{1}{2m}\,Q_j\frac{\partial}{\partial x_j} -
 \gamma^0\,\frac{1}{4m}\frac{\partial Q_j}{\partial x_j} -
 \gamma^0\,\frac{i}{4m}\,\varepsilon^{jkm}\Sigma_m \frac{\partial
   Q_k}{\partial x^j}\nonumber\\ &&+\>
 \gamma^0\,\frac{1}{m}\,\tilde{h}^0_j\frac{\partial^2}{\partial x_j
   \partial t} + \gamma^0\,\frac{1}{2m}\frac{\partial
   \tilde{h}^0_j}{\partial x_j}\frac{\partial}{\partial t} +
 \gamma^0\,\frac{i}{2m}\,\varepsilon^{jkm} \Sigma_m \frac{\partial
   \tilde{h}^0_k}{\partial x^j}\frac{\partial}{\partial t}\>.
\end{eqnarray}
Here $\Sigma_1 = -\Sigma^1 = -(\vec{\Sigma}\,)_x$ etc. and $\vec{\Sigma} = \gamma^0
\vec{\gamma} \gamma^5$ is the standard diagonal matrix with elements
$(\vec{\sigma}, \vec{\sigma}\,)$ and $\vec{\sigma} =
(\sigma_1,\sigma_2,\sigma_3)$ are the Pauli $2\times 2$ matrices
\cite{Itzykson1980}. For the Hamilton operator ${\rm H}_1$ we obtain
the following expression
\begin{eqnarray}\label{eq:58}
{\rm H}_1 &=& \gamma^0\,m - \gamma^0\,\frac{1}{2m}\,\Delta +
\gamma^0\,\tilde{h}^0_0 m +
\gamma^0\,\frac{1}{2m}\,\vec{\nabla}\tilde{h}^0_0\cdot \vec{\nabla} +
\gamma^0\,\frac{1}{8m}\,\Delta \tilde{h}^0_0 +
\gamma^0\,\frac{1}{2m}\,\tilde{h}^0_0 \Delta +
\gamma^0\,\frac{i}{4m}\,\vec{\Sigma}\cdot
(\vec{\nabla}\tilde{h}^0_0 \times \vec{\nabla}\,)
\nonumber\\ &&+\>\gamma^0\,\frac{1}{m}\, Q^k_j\frac{\partial^2}{\partial
  x^k \partial x_j} + \gamma^0\,\frac{1}{2m}\frac{\partial
  Q^k_j}{\partial x_j}\frac{\partial}{\partial x^k} +
\gamma^0\,\frac{i}{2m}\,\varepsilon^{jkm} \Sigma_m \frac{\partial
  Q^{\ell}_k}{\partial x^j}\frac{\partial}{\partial
  x^{\ell}}\nonumber\\ &&+\> \gamma^0\,\frac{1}{2m}\,Q_j
\frac{\partial}{\partial x_j} + \gamma^0\,\frac{1}{4m}\frac{\partial
  Q_j}{\partial x_j} +
\gamma^0\,\frac{i}{4m}\,\varepsilon^{jkm}\Sigma_m\frac{\partial
  Q_k}{\partial x^j}\nonumber\\ &&-\>\gamma^0\,\frac{1}{m}\, \tilde{h}^0_j
\frac{\partial^2}{\partial x_j \partial t} -
\gamma^0\,\frac{1}{2m}\frac{\partial \tilde{h}^0_j}{\partial
  x_j}\frac{\partial}{\partial t} -
\gamma^0\,\frac{i}{2m}\,\varepsilon^{jkm}\Sigma_m \frac{\partial
  \tilde{h}^0_k}{\partial x^j}\frac{\partial}{\partial
  t}\nonumber\\ &&+\> \frac{i}{2}\,\gamma^0\gamma^j\frac{\partial
  \tilde{h}^0_0}{\partial x^j} + i\,\gamma^0 \gamma^j \tilde{h}^0_0
\frac{\partial}{\partial x^j} + \frac{1}{2m}\,\gamma^j\frac{\partial
  \tilde{h}^0_k}{\partial x^j}\frac{\partial}{\partial x_k} +
\frac{1}{4m}\,\gamma^j\frac{\partial^2 \tilde{h}^0_k}{\partial x^j
  \partial x_k} - \frac{1}{2m}\, \gamma^j\frac{\partial
  \tilde{h}^0_j}{\partial t}\frac{\partial}{\partial t}\nonumber\\ &&+\>
\frac{1}{2m}\,\gamma^j\frac{\partial Q^k_j}{\partial
  t}\frac{\partial}{\partial x^j} + \frac{1}{4m}\,\gamma^j
\frac{\partial Q_j}{\partial t} + \frac{i}{2}\frac{\partial
  \tilde{h}^j_0}{\partial x^j} + i
\tilde{h}^j_0\frac{\partial}{\partial x^j}\>.
\end{eqnarray}
For the derivation of the Hamilton operator Eq.~(\ref{eq:58}) we have
neglected contributions of order $1/m^2$. Since there are still so-called {\it odd}--operators left, we have to perform another FW transformation
\begin{eqnarray}\label{eq:59}
{\rm H}_2 &=& e^{\,+ i S_2}\,{\rm H}_1\,e^{\,-i S_2} - i\,e^{\,i
  S_2}\frac{\partial}{\partial t}e^{\,-i S_2} \nonumber\\
  &=& {\rm H}_1 -
\frac{\partial S_2}{\partial t} + i\Big[S_2,{\rm H}_1 -
  \frac{1}{2}\frac{\partial S_2}{\partial t}\Big] +
\frac{i^2}{2}\,\Big[S_2,\Big[S_2,{\rm H}_1 -
    \frac{1}{3}\frac{\partial S_2}{\partial t}\Big]\Big] + \ldots\>,
\end{eqnarray}
where the operator $S_2$ is equal to
\begin{eqnarray}\label{eq:60}
S_2 &=& - \frac{i}{2m}\,\gamma^0\Big(\frac{i}{2}\,\gamma^0\gamma^j
\frac{\partial \tilde{h}^0_0}{\partial x^j} + i\,\gamma^0 \gamma^j
\tilde{h}^0_0 \frac{\partial}{\partial x^j} +
\frac{1}{2m}\,\gamma^j\frac{\partial \tilde{h}^0_k}{\partial
  x^j}\frac{\partial}{\partial x_k} +
\frac{1}{4m}\,\gamma^j\frac{\partial^2 \tilde{h}^0_k}{\partial x^j
  \partial x_k} - \frac{1}{2m}\, \gamma^j\frac{\partial
  \tilde{h}^0_j}{\partial t}\frac{\partial}{\partial t}\nonumber\\ &&+\>
\frac{1}{2m}\,\gamma^j\frac{\partial Q^k_j}{\partial
  t}\frac{\partial}{\partial x^j} + \frac{1}{4m}\,\gamma^j
\frac{\partial Q_j}{\partial t}\Big) \nonumber\\ &=&
\frac{1}{4m}\,\gamma^j\frac{\partial \tilde{h}^0_0}{\partial x^j} +
\frac{1}{2m}\, \gamma^j \tilde{h}^0_0 \frac{\partial}{\partial x^j} -
\frac{i}{4m^2}\,\gamma^0\gamma^j\frac{\partial
  \tilde{h}^0_k}{\partial x^j}\frac{\partial}{\partial x_k}
- \frac{i}{8m^2}\,\gamma^0\gamma^j\frac{\partial^2
    \tilde{h}^0_k}{\partial x^j \partial x_k} +
\frac{i}{4m^2}\,\gamma^0\gamma^j\frac{\partial
  \tilde{h}^0_j}{\partial t}\frac{\partial}{\partial
  t}\nonumber\\ &&-\>
    \frac{i}{4m^2}\,\gamma^0\gamma^j\frac{\partial Q^k_j}{\partial
      t}\frac{\partial}{\partial x^j} -
    \frac{i}{8m^2}\,\gamma^0\gamma^j\frac{\partial Q_j}{\partial t}\>.
\end{eqnarray}
Keeping only contributions of order $1/m$ for the time derivative
of $S_2$ and the commutators we obtain the following expressions
\begin{eqnarray}\label{eq:61}
\frac{\partial S_2}{\partial t} &=& \frac{1}{4m}\,\vec{\gamma}\cdot
\vec{\nabla}\frac{\partial \tilde{h}^0_0}{\partial t} +
\frac{1}{2m}\frac{\partial \tilde{h}^0_0}{\partial
  t}\,\vec{\gamma}\cdot \vec{\nabla}\>,\nonumber\\ i\Big[S_2,{\rm H}_1 -
  \frac{1}{2}\frac{\partial S_2}{\partial t}\Big]&=& -
\frac{i}{2}\,\gamma^0\gamma^j\frac{\partial \tilde{h}^0_0}{\partial
  x^j} - i\,\gamma^0 \gamma^j \tilde{h}^0_0 \frac{\partial}{\partial
  x^j} - \frac{1}{2m}\,\gamma^j\frac{\partial
  \tilde{h}^0_k}{\partial x^j}\frac{\partial}{\partial x_k} -
\frac{1}{4m}\,\gamma^j\frac{\partial^2 \tilde{h}^0_k}{\partial x^j
  \partial x_k} + \frac{1}{2m}\, \gamma^j\frac{\partial
  \tilde{h}^0_j}{\partial t}\frac{\partial}{\partial
  t}\nonumber\\ &&-\> \frac{1}{2m}\,\gamma^j\frac{\partial
    Q^k_j}{\partial t}\frac{\partial}{\partial x^j} -
  \frac{1}{4m}\,\gamma^j\frac{\partial Q_j}{\partial
    t}\>,\nonumber\\ \frac{i^2}{2}\,\Big[S_2,\Big[S_2,{\rm H}_1 -
        \frac{1}{3}\frac{\partial S_2}{\partial t}\Big]\Big] &=& 0\>.
\end{eqnarray}
Hence, after two FW transformations the effective Hamilton operator
takes the form
\begin{eqnarray}\label{eq:62}
{\rm H}_2 &=& \gamma^0\,m - \gamma^0\,\frac{1}{2m}\,\Delta +
\gamma^0\,\tilde{h}^0_0 m +
\gamma^0\,\frac{1}{2m}\,\vec{\nabla}\tilde{h}^0_0\cdot
  \vec{\nabla} + \gamma^0\,\frac{1}{8m}\,\Delta \tilde{h}^0_0 +
  \gamma^0\,\frac{1}{2m}\,\tilde{h}^0_0 \Delta +
  \gamma^0\,\frac{i}{4m}\,\vec{\Sigma}\cdot (\vec{\nabla}\tilde{h}^0_0
  \times \vec{\nabla}\,)\nonumber\\ &&+\>\gamma^0\,\frac{1}{m}\,
Q^k_j\frac{\partial^2}{\partial x^k \partial x_j} +
\gamma^0\,\frac{1}{2m}\frac{\partial Q^k_j}{\partial
  x_j}\frac{\partial}{\partial x^k} +
\gamma^0\,\frac{i}{2m}\,\varepsilon^{jkm} \Sigma_m \frac{\partial
  Q^{\ell}_k}{\partial x^j}\frac{\partial}{\partial
  x^{\ell}}\nonumber\\ &&+\> \gamma^0\,\frac{1}{2m}\,Q_j
\frac{\partial}{\partial x_j} + \gamma^0\,\frac{1}{4m}\frac{\partial
  Q_j}{\partial x_j} +
\gamma^0\,\frac{i}{4m}\,\varepsilon^{jkm}\Sigma_m\frac{\partial
  Q_k}{\partial x^j}\nonumber\\ &&-\>\gamma^0
  \frac{1}{m}\, \tilde{h}^0_j \frac{\partial^2}{\partial x_j \partial t}
  - \gamma^0\,\frac{1}{2m}\frac{\partial \tilde{h}^0_j}{\partial
    x_j}\frac{\partial}{\partial t} -
  \gamma^0\,\frac{i}{2m}\,\varepsilon^{jkm}\Sigma_m \frac{\partial
    \tilde{h}^0_k}{\partial x^j}\frac{\partial}{\partial
    t}\nonumber\\ &&-\> \frac{1}{4m}\,\vec{\gamma}\cdot
  \vec{\nabla}\frac{\partial \tilde{h}^0_0}{\partial t} -
  \frac{1}{2m}\frac{\partial \tilde{h}^0_0}{\partial
    t}\,\vec{\gamma}\cdot \vec{\nabla}  + \frac{i}{2}\frac{\partial
   \tilde{h}^j_0}{\partial x^j} + i
   \tilde{h}^j_0\frac{\partial}{\partial x^j}\>.
\end{eqnarray}
The remaining {\it odd}--operators may be deleted by the final FW transformation
\begin{eqnarray}\label{eq:63}
{\rm H}_3 &=& e^{\,+ i S_3}\,{\rm H}_2\,e^{\,-i S_3} - i\,e^{\,i
  S_3}\frac{\partial}{\partial t}e^{\,-i S_3} \nonumber\\
  &=& {\rm H}_2 -
\frac{\partial S_3}{\partial t} + i\Big[S_3,{\rm H}_2 -
  \frac{1}{2}\frac{\partial S_3}{\partial t}\Big] +
\frac{i^2}{2}\,\Big[S_3,\Big[S_3,{\rm H}_2 -
    \frac{1}{3}\frac{\partial S_3}{\partial t}\Big]\Big] + \ldots\>,
\end{eqnarray}
with the operator $S_3$ given by
\begin{eqnarray}\label{eq:64}
S_3 &=& - \frac{i}{2m}\,\gamma^0\,\Big( -
\frac{1}{4m}\,\vec{\gamma}\cdot \vec{\nabla}\frac{\partial
  \tilde{h}^0_0}{\partial t} -
  \frac{1}{2m}\frac{\partial \tilde{h}^0_0}{\partial
      t}\,\vec{\gamma}\cdot \vec{\nabla}\Big) \nonumber\\
      &=& 
    \frac{1}{8m^2}\,i\,\gamma^0 \vec{\gamma}\cdot
    \vec{\nabla}\frac{\partial \tilde{h}^0_0}{\partial t} +
  \frac{1}{4m^2}\frac{\partial \tilde{h}^0_0}{\partial
      t}\,i\,\gamma^0 \vec{\gamma}\cdot \vec{\nabla}\>.
\end{eqnarray}
Again, keeping only contributions of order $1/m$ for the time derivative
of $S_3$ and the commutators we obtain the following expressions
\begin{eqnarray}\label{eq:65}
\frac{\partial S_3}{\partial t} &=& 0\>,\nonumber\\ i\Big[S_3,{\rm
    H}_2 - \frac{1}{2}\frac{\partial S_3}{\partial
    t}\Big]&=&\frac{1}{4m}\,\vec{\gamma}\cdot
  \vec{\nabla}\frac{\partial \tilde{h}^0_0}{\partial t} +
  \frac{1}{2m}\frac{\partial \tilde{h}^0_0}{\partial
    t}\,\vec{\gamma}\cdot \vec{\nabla}\>,\nonumber\\ \frac{i^2}{2}\,\Big[S_3,\Big[S_3,{\rm H}_2 -
    \frac{1}{3}\frac{\partial S_3}{\partial t}\Big]\Big]&=& 0\>.
\end{eqnarray}
Finally, we obtain
the low--energy reduction of the Dirac Hamilton operator for slow
fermions 
\begin{eqnarray}\label{eq:66}
{\rm H}_3 &=& \gamma^0\,m - \gamma^0\,\frac{1}{2m}\,\Delta +
\gamma^0\,\tilde{h}^0_0 m +
\gamma^0\,\frac{1}{2m}\,\vec{\nabla}\tilde{h}^0_0\cdot \vec{\nabla} +
\gamma^0\,\frac{1}{8m}\,\Delta \tilde{h}^0_0 +
\gamma^0\,\frac{1}{2m}\,\tilde{h}^0_0 \Delta +
\gamma^0\,\frac{i}{4m}\,\vec{\Sigma}\cdot
(\vec{\nabla}\tilde{h}^0_0 \times \vec{\nabla}\,)
\nonumber\\ &&+\>\gamma^0\,\frac{1}{m}\, Q^k_j\frac{\partial^2}{\partial
  x^k \partial x_j} + \gamma^0\,\frac{1}{2m}\frac{\partial
  Q^k_j}{\partial x_j}\frac{\partial}{\partial x^k} +
\gamma^0\,\frac{i}{2m}\,\varepsilon^{jkm} \Sigma_m \frac{\partial
  Q^{\ell}_k}{\partial x^j}\frac{\partial}{\partial
  x^{\ell}}\nonumber\\ &&+\> \gamma^0\,\frac{1}{2m}\,Q_j
\frac{\partial}{\partial x_j} + \gamma^0\,\frac{1}{4m}\frac{\partial
  Q_j}{\partial x_j} +
\gamma^0\,\frac{i}{4m}\,\varepsilon^{jkm}\Sigma_m\frac{\partial
  Q_k}{\partial x^j}\nonumber\\ &&-\>\gamma^0\,\frac{1}{m}\, \tilde{h}^0_j
\frac{\partial^2}{\partial x_j \partial t} -
\gamma^0\,\frac{1}{2m}\frac{\partial \tilde{h}^0_j}{\partial
  x_j}\frac{\partial}{\partial t} -
\gamma^0\,\frac{i}{2m}\,\varepsilon^{jkm}\Sigma_m \frac{\partial
  \tilde{h}^0_k}{\partial x^j}\frac{\partial}{\partial t} +
\frac{i}{2}\frac{\partial \tilde{h}^j_0}{\partial x^j} + i
\tilde{h}^j_0\frac{\partial}{\partial x^j}\>.
\end{eqnarray}
Following the standard procedure \cite{Foldy1950} (see also
\cite{Ivanov2014} for further details), we skip
intermediate calculations and arrive at the Schr\"odinger--Pauli
equation for the large components of the Dirac wave function of slow
fermions
\begin{eqnarray}\label{eq:67}
i\,\frac{\partial \Psi(\vec{r},t)}{\partial t} = \Big( -
\frac{1}{2m}\,\Delta + \Phi_{\rm mgr}\Big)\,\Psi(\vec{r},t)\>,
\end{eqnarray}
where $\Psi(t, \vec{r}\,)$ is the large component of the Dirac wave
function and $\Phi_{\rm mgr}$ the effective
low--energy potential caused by massive gravity
\begin{eqnarray}\label{eq:68}
\Phi_{\rm mgr} &=&\tilde{h}^0_0 m
-\frac{1}{2m}\,\vec{\nabla}\tilde{h}^0_0 \cdot \vec{\nabla} -
\frac{1}{8m}\,\Delta \tilde{h}^0_0 - \frac{1}{2m}\,\tilde{h}^0_0
\Delta - \frac{i}{4m}\,\vec{\sigma}\cdot (\vec{\nabla}\tilde{h}^0_0
\times \vec{\nabla}\,) \nonumber\\ &&-\> \frac{1}{m}\,\tilde{h}^k_j
\frac{\partial^2}{\partial x^k \partial x_j} -
\frac{1}{m}\frac{\partial \tilde{h}^k_j}{\partial
  x_j}\frac{\partial}{\partial x^k} - \frac{1}{4m}\frac{\partial^2
  \tilde{h}^k_j}{\partial x_j \partial x^k} -
\frac{i}{2m}\,\varepsilon^{jkm}\sigma_m\,\frac{\partial
  \tilde{h}^{\ell}_k}{\partial x^j}\frac{\partial}{\partial
  x^{\ell}} -
\frac{i}{4m}\,\varepsilon^{jkm}\sigma_m\,\frac{\partial^2
  \tilde{h}^{\ell}_k}{\partial x^j \partial x^{\ell}}\nonumber\\ &&+\>
i\frac{\partial \tilde{h}^0_j}{\partial x_j} + 2i
\tilde{h}^0_j\frac{\partial}{\partial x_j} -
\frac{1}{2m}\frac{\partial \tilde{h}^0_j}{\partial
  t}\frac{\partial}{\partial x_j} - \frac{1}{4m}\frac{\partial^2
  \tilde{h}^0_j}{\partial x_j \partial t} - \frac{1}{m}\, \tilde{h}^0_j
\frac{\partial^2}{\partial x_j \partial t} -
\frac{1}{2m}\frac{\partial \tilde{h}^0_j}{\partial
  x_j}\frac{\partial}{\partial t} \nonumber\\ &&- \frac{1}{2}\,\varepsilon^{jkm}
\sigma_m \frac{\partial \tilde{h}^0_k}{\partial x^j} -\>
\frac{i}{2m}\,\varepsilon^{jkm}\sigma_m \frac{\partial
  \tilde{h}^0_k}{\partial x^j}\frac{\partial}{\partial t} -
\frac{i}{4m}\,\varepsilon^{jkm}\sigma_m\,\frac{\partial^2
  \tilde{h}^0_k}{\partial x^j \partial t}\>.
\end{eqnarray}
Above, we have made use of $\sigma_1 = -\sigma^1 =
-(\vec{\sigma}\,)_x$ etc.  and have used for $Q^k_j$ and $Q_j$ the
corresponding expressions in terms of $\tilde{h}^{\mu}_{\nu}$ in
Eq.~(\ref{eq:51}). Some terms of order $\mathcal O(1)$
  in the large $m$--expansion appear in the effective low--energy
  potential Eq.~(\ref{eq:68}) due to the removal of the mass term in
  the Hamilton operator Eq.~(\ref{eq:66}) by means of the fermion wave
  function transformation with the phase factor $e^{\,-i m t}$ (see
  \cite{Ivanov2014} for details). We would like to note that setting
$\tilde{h}^0_0 = U$, $\tilde{h}^k_j = - \gamma\,U\,\delta^k_j$ and
$\tilde{h}^0_j = 0$ we arrive at the effective gravitational potential
given by Eq.~(\ref{eq:20}) of Ref.~\cite{Ivanov2014}.

\section{Massive graviton, scalar and vector fields caused by
massive point--like body with mass $M$}
\label{sec:compare}

Here we would like to compare our results with those
  obtained by Gon\c{c}alves, Obukhov, and Shapiro \cite{Obukhov2007}
  as well as Quach \cite{Quach2015}. The main distinction lies in the
  fact that the results, obtained by these authors, are not related to
  linearised massive gravity but instead are derived for massless
  gravitons. The effective low--energy Hamilton operator
  Eq.~(\ref{eq:37}) of Ref.~\cite{Obukhov2007} is derived for
  gravitational fields  (gravitational waves) in
  spacetimes with metric
\begin{eqnarray}\label{eq:69}
 ds^2 = - dt^2 + dx^2 + (1 - 2 v)\,dy^2 + (1 + 2 v)\,dz^2 - 2u\,dydz -
 2u\,dz dy,
\end{eqnarray}
for $u = 0$ and $v = v(t - x)$. Apart from the different signature the
metric Eq.~(\ref{eq:69}) is, of course, a partial case of the metric
used for the derivation of the effective low--energy potential
Eq.~(\ref{eq:68}). In Ref.~\cite{Quach2015} the metric
Eq.~(\ref{eq:69}) with the replacement $v \to f$ has been used for the
derivation of the effective low--energy Hamilton operator
Eq.~(\ref{eq:21}). Switching off the electromagnetic field, taken into
account in Refs.~\cite{Obukhov2007,Quach2015}, the effective
low--energy potential $\Phi_{\rm eff}$, which can be obtained from
Eq.~(\ref{eq:37}) of Ref.~\cite{Obukhov2007} and Eq.~(\ref{eq:21}) of
Ref.~\cite{Quach2015}, is given by
\begin{eqnarray}\label{eq:70}
 \Phi_{\rm eff} = - \frac{1}{m}\,v\,T^{ab}\,\frac{\partial^2}{\partial
   x^a \partial x^b} -
 \frac{i}{2m}\,\epsilon^{abc}\,\sigma_a\frac{\partial v}{\partial
   x^b}\,{T^d}_c\,\frac{\partial}{\partial x^d}\>,
\end{eqnarray}
where $T = {\rm diag}(0, -1, + 1)$ \cite{Obukhov2007,Quach2015}. The
potential $\Phi_{\rm eff}$ does not reproduce the effective
low--energy potential $\Phi_{\rm mgr}$ given in Eq.~(\ref{eq:68}). It
can be compared only with the part of Eq.~(\ref{eq:68}) equal to
\begin{eqnarray}\label{eq:71}
 \delta \Phi_{\rm mgr} = - \frac{1}{m}\,\tilde{h}^k_j
 \frac{\partial^2}{\partial x^k \partial x_j} -
 \frac{i}{2m}\,\varepsilon^{jkm}\sigma_m\,\frac{\partial
 \tilde{h}^{\ell}_k}{\partial x^j}\frac{\partial}{\partial x^{\ell}}
\end{eqnarray}
for a corresponding choice of $\tilde{h}^k_j$.
Because of a specific choice of the metric
  Eq.~(\ref{eq:69}) with functions $u = 0$ and $v = v(t - x)$,
  describing a gravitational wave propagating along $x$ axis and
  having one polarization state \cite{Obukhov2007}, the results in
\cite{Obukhov2007} and \cite{Quach2015}, cannot be applied to
experimental investigations of linearised massive gravity induced by
massive bodies for slow Dirac fermions, in contrast to our results
which allow for that.  As an example, we discuss the gravitational
field of a point--like mass $M$. Following \cite{Hinterbichler2012} we
take as source a point--like mass $M$ with energy--momentum tensor
\begin{eqnarray}\label{eq:72}
 T^{(M)}_{\mu\nu} = M\,\delta^0_{\mu}\delta^0_{\nu}\,\delta^{(3)}(\vec{r}\,),
\end{eqnarray}
where $\delta^{(3)}(\vec{r}\,)$ is the Dirac $\delta$--function.  For
the experimental analysis we consider static solutions of the equations of motion, which are given by \cite{Hinterbichler2012} 
\begin{eqnarray}\label{eq:73}
 (\Delta - m^2_g)\,h^{(M)}_{\mu\nu}(\vec{r}\,) &=& \frac{M}{M_{\rm
      Pl}}\,\Big(\delta^0_{\mu}\delta^0_{\nu} - \frac{1}{2}\,\eta_{\mu\nu}\Big)\,\delta^{(3)}(\vec{r}\,),\nonumber\\ 
      (\Delta - m^2_g)\,A^{(M)}_{\alpha}(\vec{r}\,) &=& 0, \nonumber\\
      (\Delta - m^2_g)\,\varphi^{(M)}(\vec{r}\,) &=& d\,\frac{M}{M_{\rm
      Pl}}\,\delta^{(3)}(\vec{r}\,),
\end{eqnarray}
where we have used $6d^2 = 1$. Taking into account the gauge conditions
\cite{Hinterbichler2012,Nibbelink2007}
\begin{eqnarray}\label{eq:74}
 \frac{\partial h^{(M)}_{\nu\alpha}}{\partial x_{\nu}} -
 \frac{1}{2}\frac{\partial h^{(M)}}{\partial x^{\alpha}} -
 \frac{1}{\sqrt{2}}\,m_g\,A^{(M)}_{\alpha} &=& 0,\nonumber\\ \frac{\partial
 A^{(M)}_{\mu}}{\partial x_{\mu}} + \frac{1}{\sqrt{2}}\,m_g\,h^{(M)} + 3 d
 \sqrt{2}\,m_g\,\varphi^{(M)} &=& 0,
\end{eqnarray}
we adduce the solutions
\begin{eqnarray}\label{eq:75}
      h^{(M)}_{\mu\nu}(\vec{r}\,) &=& -\frac{M}{M_{\rm
      Pl}}\,\Big(\delta^0_{\mu}\delta^0_{\nu} - \frac{1}{2}\,\eta_{\mu\nu}\Big)\,\frac{1}{4\pi}\frac{e^{- m_g r}}{r},\nonumber\\ 
      A^{(M)}_{\alpha}(\vec{r}\,) &=& 0, \nonumber\\
      \varphi^{(M)}(\vec{r}\,) &=& -d\,\frac{M}{M_{\rm
      Pl}}\frac{1}{4\pi}\frac{e^{- m_g r}}{r}.
\end{eqnarray}
The massive gravitational field $\tilde{h}^{(M)}_{\mu\nu}$, which couples to slow neutrons, is equal to
\begin{align}\label{eq:76}
\tilde{h}^{(M)}_{\mu\nu} = -\frac{M}{M_\text{Pl}^2}\,\bigg\{ \delta_\mu^0\delta_\nu^0 - \frac{1}{3}\,\Big(\eta_{\mu\nu} + \frac{1}{m_g^2}\frac{\partial^2}{\partial x^\mu\partial
 x^\nu}\Big)\bigg\}\,\frac{1}{4\pi}\frac{e^{-m_gr}}{r},
\end{align}
and obeys the constraint $\tilde{h}^{(M)} = 0$. One can see that the
gravitational field $\tilde{h}^{(M)}_{ij}$ becomes singular in the
limit of a vanishing graviton mass $m_g \to 0$. Following
Arkani--Hamed, Georgi, and Schwartz \cite{Georgi2003} (see also
\cite{Vainshtein2006}) such a singularity is due to the restricted
applicability of linearised massive gravity only for distances much
larger than the Vainshtein radius $r_V = \sqrt[3]{M/m^2_g M^2_{\rm
    Pl}}$, i.e. $r \gg r_V = \sqrt[3]{M/m^2_g M^2_{\rm Pl}}$. In terms
of the Vainshtein radius the gravitational field
$\tilde{h}^{(M)}_{\mu\nu}$ reads
\begin{align}\label{eq:77}
\tilde{h}^{(M)}_{\mu\nu} = -r^3_V\,\bigg\{m_g^2\,\delta_\mu^0\delta_\nu^0 - \frac{1}{3}\,\Big(m_g^2\,\eta_{\mu\nu} + \frac{\partial^2}{\partial x^\mu\partial
 x^\nu}\Big)\bigg\}\,\frac{1}{4\pi}\frac{e^{-m_gr}}{r}.
\end{align}
Clearly, such an analysis of static gravitational fields generated by
massive bodies in terms of their gravitational interactions with slow
fermions is not possible by using the results, obtained in
\cite{Obukhov2007,Quach2015}.

Following \cite{Obukhov2007,Quach2015} and using the effective
low--energy potential Eq.~(\ref{eq:68}) we calculate the neutron spin
precession in the gravitational field within linearised massive
gravity
\begin{eqnarray}\label{eq:78}
 \frac{d\vec{S}}{dt} = \vec{\Omega}_{\rm mgr} \times \vec{S},
\end{eqnarray}
where $\vec{S} = \frac{1}{2}\,\vec{\sigma}$ is the 
neutron spin operator and $\vec{\Omega}_{\rm mgr}$ is the angular velocity
operator of the neutron spin precession equal to
\begin{eqnarray}\label{eq:79}
 \Omega^a_{\rm mgr} = \varepsilon^{ajk} \bigg(\frac{\partial
 \tilde{h}^0_k}{\partial x^j} - \frac{i}{2m}\frac{\partial
 \tilde{h}^0_0}{\partial x^j}\frac{\partial}{\partial x^k} +
 \frac{i}{m}\frac{\partial \tilde{h}^0_k}{\partial
 x^j}\frac{\partial}{\partial t} + \frac{i}{2m}\frac{\partial^2
 \tilde{h}^0_k}{\partial x^j \partial t} +
 \frac{i}{m}\frac{\partial \tilde{h}^{\ell}_k}{\partial
 x^j}\frac{\partial}{\partial x^{\ell}} +
 \frac{i}{2m}\frac{\partial^2 \tilde{h}^{\ell}_k}{\partial x^j
 \partial x^{\ell}}\bigg).
\end{eqnarray}
The phase-shift of the neutron wave function, induced by the effective
low--energy operator $\Phi_{\rm spin} = \vec{\Omega}_{\rm mgr} \cdot
\vec{S}$ can in principle be measured by neutron interferometers
\cite{Rauch2015, Rauch1974}--\cite{Demirel2015}.
Using Eq.~(\ref{eq:70}) we may calculate the angular
  velocity operator of the neutron spin precession
\begin{eqnarray}\label{eq:80}
 \Omega^a_{\rm gr} = \epsilon^{abc}\,\frac{i}{m}\frac{\partial
 v}{\partial x^b}\,{T^d}_c\,\frac{\partial}{\partial x^d},
\end{eqnarray}
which can be compared with only one term in the angular velocity
operator $\Omega^a_{\rm mgr}$ in Eq.~(\ref{eq:79}).

Thus, apart from non--vanishing electromagnetic field contributions,
which are taken into account in \cite{Obukhov2007,Quach2015}, we may
argue that the effective low--energy gravitational potential given in
Eq.~(\ref{eq:68}) is a generalization of the corresponding results in
\cite{Obukhov2007} and \cite{Quach2015}.

\section{Conclusion}
\label{sec:conclusion}

We have analysed the Dirac equation for slow fermions within
linearised massive gravity \cite{Hinterbichler2012,Khoury2015}. We
treat massive gravity as a linear expansion above the Minkowski
background. According to \cite{Georgi2003} (see also
\cite{Vainshtein2006}), such a theory is applicable to observable
phenomena at distances much larger than the Vainshtein radius $r \gg
r_V$, where $r_V = \sqrt[3]{M/m^2_g M^2_{\rm Pl}}$ and $M$ is a
gravitating mass. We would like to emphasize that unlike Vainshtein
\cite{Vainshtein2006}, who claimed that the radius of applicability of
linearised massive gravity is $r_V = \sqrt[5]{M/m^4_g M^2_{\rm Pl}}$,
we have found that the radius of applicability is $r_V =
\sqrt[3]{M/m^2_g M^2_{\rm Pl}}$ instead, which is in complete
agreement with the analysis by Arkani--Hamed, Georgi, and Schwartz
\cite{Georgi2003}.

Using Foldy--Wouthuysen transformations we have derived the effective
low--energy gravitational potential for slow fermions coupled to the
fields of linearised massive gravity.  We have used a version of
linearised massive gravity employing St\"uckelberg tricks in which the
Fierz--Pauli (FP) massive graviton $\tilde{h}_{\mu\nu}$ with mass
$m_g$ is decomposed into a linear superposition of a tensor
$h_{\mu\nu}$, vector $A_{\mu}$ and scalar field $\varphi$.  The slow
fermions couple to these fields only in terms of the FP field
$\tilde{h}_{\mu\nu}$, which are functions of time and spatial
coordinates. This implies that the results obtained in this paper can
be used for the analysis of the interaction of slow fermions in the
terrestrial laboratories coupled to gravitational waves, which are
emitted by cosmological objects, within the context of massive gravity
\cite{Brito2013}. This might allow to understand the role of massive
gravitons in the dynamics of the evolution of the Universe
\cite{Eckhardt2010,D'Amico2011,Felice2013,Camelli2014} and
instabilities of black holes \cite{Babichev2004}-\cite{Babichev2013}.

Finally, we have compared our results with the results, obtained by
Gon\c{c}alves, Obukhov, and Shapiro \cite{Obukhov2007} and as well as
by Quach \cite{Quach2015}. We have shown that the effective
low--energy potentials, calculated in \cite{Obukhov2007,Quach2015},
are obtained for massless gravitons and, hence, have no relation to
linearised massive gravity.  These potentials, obtained as functionals
of the metric Eq.~(\ref{eq:69}) for $u = 0$ and $v = v(t - x)$,
describing a gravitational wave with one polarization state and
propagating along $x$ axis, cannot describe interactions of static
gravitational fields of massive bodies and their effect on slow Dirac
fermions.  Since a massive graviton has five polarization states
\cite{Hinterbichler2012}, one might assert that the potentials,
derived in \cite{Obukhov2007,Quach2015}, can only partly account for
the effects, which might be analysed in our approach. Hence, one may
conclude that the effective low--energy potential Eq.~(\ref{eq:68}) is
a generalization to the effective low--energy potentials calculated in
\cite{Obukhov2007,Quach2015}.
  
\section{Acknowledgements}

We thank Hartmut Abele and Justin Khoury for stimulating
discussions. This work was supported by the Austrian ``Fonds zur
F\"orderung der Wissenschaftlichen Forschung'' (FWF) under contract
I689-N16.

\end{document}